\documentclass[aps,prb,twocolumn,groupedaddress,showpacs,superscriptaddress]{revtex4-1}
\usepackage{graphicx,amsmath}
\usepackage{amssymb}
\usepackage{xcolor}
\DeclareUnicodeCharacter{200B}{}
\DeclareUnicodeCharacter{202F}{ }

\begin{document}
\title{Graph-based block-diagonalization of full configuration interaction Hamiltonian: H$_2$ chains study}
\author{Hayun Park}
\affiliation{Department of Liberal Studies, Kangwon National University, Samcheok, 25913, Republic of Korea}
\author{Hunpyo Lee}
\email{Email: hplee@kangwon.ac.kr}
\affiliation{Department of Liberal Studies, Kangwon National University, Samcheok, 25913, Republic of Korea}
\affiliation{Quantum Sub Inc., Samcheok, 25913, Republic of Korea}
\date{\today}

\begin{abstract}
We developed a graph-based block-diagonalization (GBBD) method for the full configuration interaction 
Hamiltonian of molecular systems to efficiently calculate the exact eigenvalues of low-energy states. In this 
approach, the 
non-zero matrix elements of the Hamiltonian are represented as edges on a graph, which naturally decomposes into 
disconnected clusters. Each cluster corresponds to an independent block in the block-diagonalized form of the 
Hamiltonian. The eigenvalues in the low-energy sector were obtained by solving the eigenvalue problem for each block 
matrix and by solving a modified Hamiltonian subject to orthonormality constraints with respect to previously 
computed lower-energy eigenstates. We applied the GBBD method to linear hydrogen H chains ranging from H$_2$ to 
H$_{12}$. The results showed excellent agreement with exact ones, confirming both the accuracy and efficiency of the 
proposed method. Finally, we discussed several physical properties with respect to the number of H$_2$ molecules. 
\end{abstract}

\maketitle

\section{Introduction\label{Introduction}}

Strong electron correlations between electrons in molecular systems play a crucial role in determining 
their physical properties and chemical reactions. To accurately capture these effects beyond mean field 
approximations, the \textit{ab initio} Hamiltonian is expressed as an $L \times L$ matrix, where $L = 2^n$ 
represents the dimension of the Hilbert space and $n$ denotes the number of molecular orbitals~\cite{Szabo1996}. As 
the number of orbitals increases in complex molecules, the size of the Hamiltonian matrix grows 
exponentially, making exact diagonalization computationally infeasible~\cite{Weisse2008,Knowles1984}. To address this 
challenge, various numerical techniques such as the Variational Quantum Eigensolver (VQE), extension of VQE, Density 
Matrix Renormalization Group (DMRG), and Quantum Monte Carlo (QMC) have been developed on both quantum and 
classical computing platforms~\cite{Peruzzo2014, Arute2020,Kandala2017, McArdle2020, White1992, White1993, 
Qunova2025, Chan2011, Baiardi2020, Foulkes2001, Ceperley1980, Booth2009,Austin2012, Bauer2020}. However, these 
methods inherently involve certain approximations and may introduce specific biases. In addition, they are generally 
limited in accessing the excited-state energies or high-energy spectrum features.

The full configuration interaction (FCI) approach is considered as the most accurate and reliable method 
for predicting molecular properties and chemical reactions, as it rigorously accounts for many-body 
effects arising from electron configurations~\cite{Bartlett2007,Sherrill1999}. It directly operates on the 
Hamiltonian without any bias or approximation appeared in numerical methods. The eigenvalues $E_N$ in energy states 
are computed from the Hamiltonian using various eigensolvers, where $N$ is the energy level. 
Furthermore, the FCI serves as a valuable benchmark for assessing the validity and accuracy of 
results obtained from numerical methods such as VQE, extension of VQE, DMRG, and QMC, which may introduce errors or 
biases. However, the FCI method is significantly limited by its high computational cost and the substantial memory 
requirements needed to store the Hamiltonian, especially for large and complex molecular systems. As a result, 
ongoing efforts are directed toward improving its efficiency in terms of both computation and memory usage.

The FCI approach involves two major computational bottlenecks. (i) The first challenge lies in 
constructing the Hamiltonian from the one-body and two-body integrals of molecular systems, which encode the 
strong interactions between electron configurations. This process requires $L \times L$ operations of a classical 
computer. In addition, for large and complex molecular systems, the size of the Hamiltonian ($L \times L$, where $L 
= 2^n$) becomes prohibitively large, leading to significant memory storage issues as it may exceed the 
capacity available for storing the Hamiltonian. (ii) The second challenge is solving for the eigenvalues 
and eigenstates of the Hamiltonian. The exact diagonalization for computation of all eigenvalues in the spectrum 
typically require computational effort on the order of $L^3$, making them impractical for large $L$.

In this work, we proposed a graph-based block-diagonalization (GBBD) method of the FCI Hamiltonian to efficiently 
compute $E_N$. The molecular Hamiltonian were constructed using  the 
OpenFermion package, where the Hamiltonian is defined in Fock space with a matrix size of $L=2^n$~\cite{McClean2017}. 
Notably, the Hamiltonian is extremely sparse, with most matrix elements being zero.  We reduced the Hamiltonian to a 
subspace with 
a fixed number $m$ of electrons, which is sufficient to describe the low-energy sector of the system without 
sacrificing essential physical information. Here, the dimension of the reduced Hamiltonian is $n \choose m$. To 
optimize memory usage, we stored only the indices and values of non-zero elements of the reduced Hamiltonian. These 
non-zero elements are then represented as edges in a graph, where basis states correspond to vertices. Interestingly, 
the resulting graph decomposes into several disconnected clusters, each corresponding to an independent block in the 
block-diagonalized 
form of the reduced Hamiltonian. Since each cluster is significantly smaller than the reduced Hamiltonian, it can be 
treated independently and efficiently. We computed the eigenvalues and eigenstates of each cluster using the Quantum 
Eigensolver (QE) method on the D-Wave quantum computer recently developed by the authors~\cite{Park2024}. The QE 
method exhibits linear scaling with respect to $L$ for computation of $E_0$, enabling efficient treatment of large 
systems. 
Ultimately, we found that the eigenvalues obtained from these independent clusters, as well as from the modified 
Hamiltonian subject to orthonormality constraints with respect to previously 
computed lower-energy eigenstates in the independent block, accurately reproduced $E_N$, thereby validating the 
proposed GBBD method.

\begin{figure}
\includegraphics[width=1.0\columnwidth]{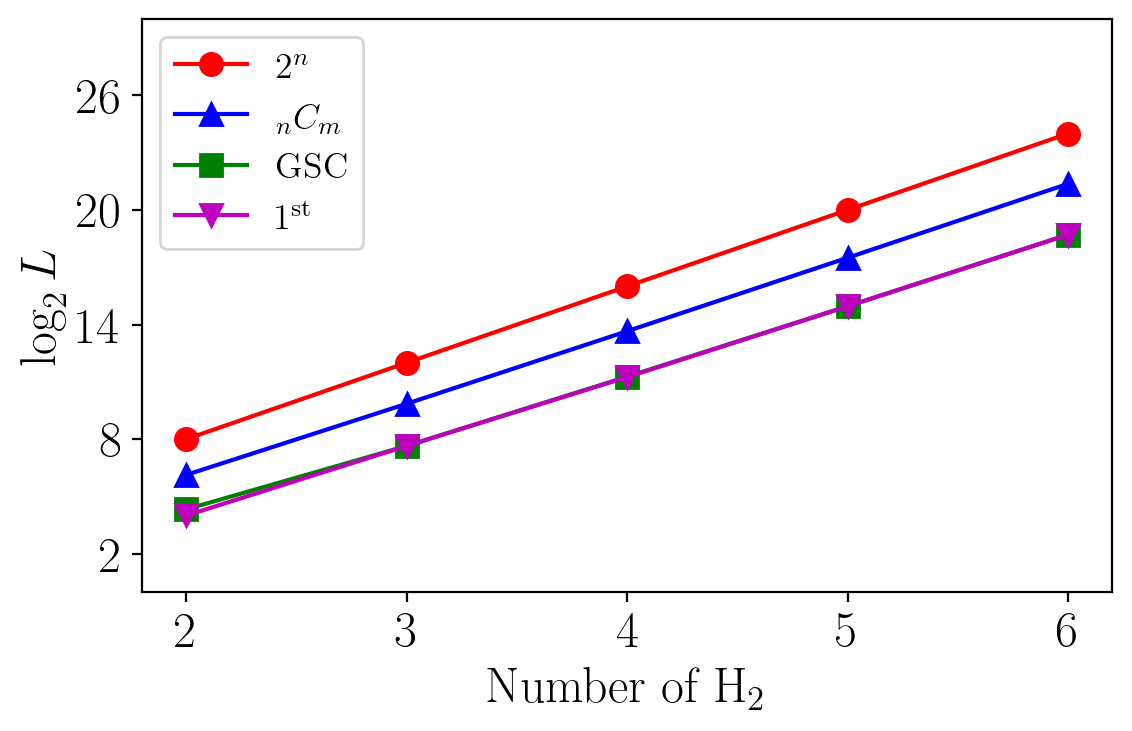}
\caption {\label{Fig1} (Color online) Logarithmic sizes of the full Hamiltonian, the reduced Hamiltonian, and the 
block matrices corresponding to the ground-state and first excited-state energies are shown as functions of the 
number of H$_2$ molecule in one-dimensional linear H$_2$ chains. The sizes of the full and reduced Hamiltonian 
correspond to 
$2^n$ and $n \choose m$, respectively, where $n$ and $m$ denote the numbers of orbitals and electrons.
‘GSC’ and ‘1$^{\text{st}}$’ indicate the ground-state and first excited-state configurations, respectively.}
\end{figure}

The main strength of the GBBD method is its capability to accurately compute eigenvalues $E_N$ for molecular 
Hamiltonian of practical size, with greatly reduced computational and memory requirements. To evaluate the 
practicality and performance of our approach, we applied the GBBD method to linear hydrogen chains ranging from H$_2$ 
to H$_{12}$ using a standard MacBook M4 Pro laptop. As a benchmark, we compared $E_N$ obtained using the GBBD method 
with the exact results computed from the reduced Hamiltonian of the H chains using the SciPy package. The comparison 
confirms that the GBBD method provides accurate results while maintaining high computational efficiency, even on 
lightweight hardware.

\begin{figure}
\includegraphics[width=1.0\columnwidth]{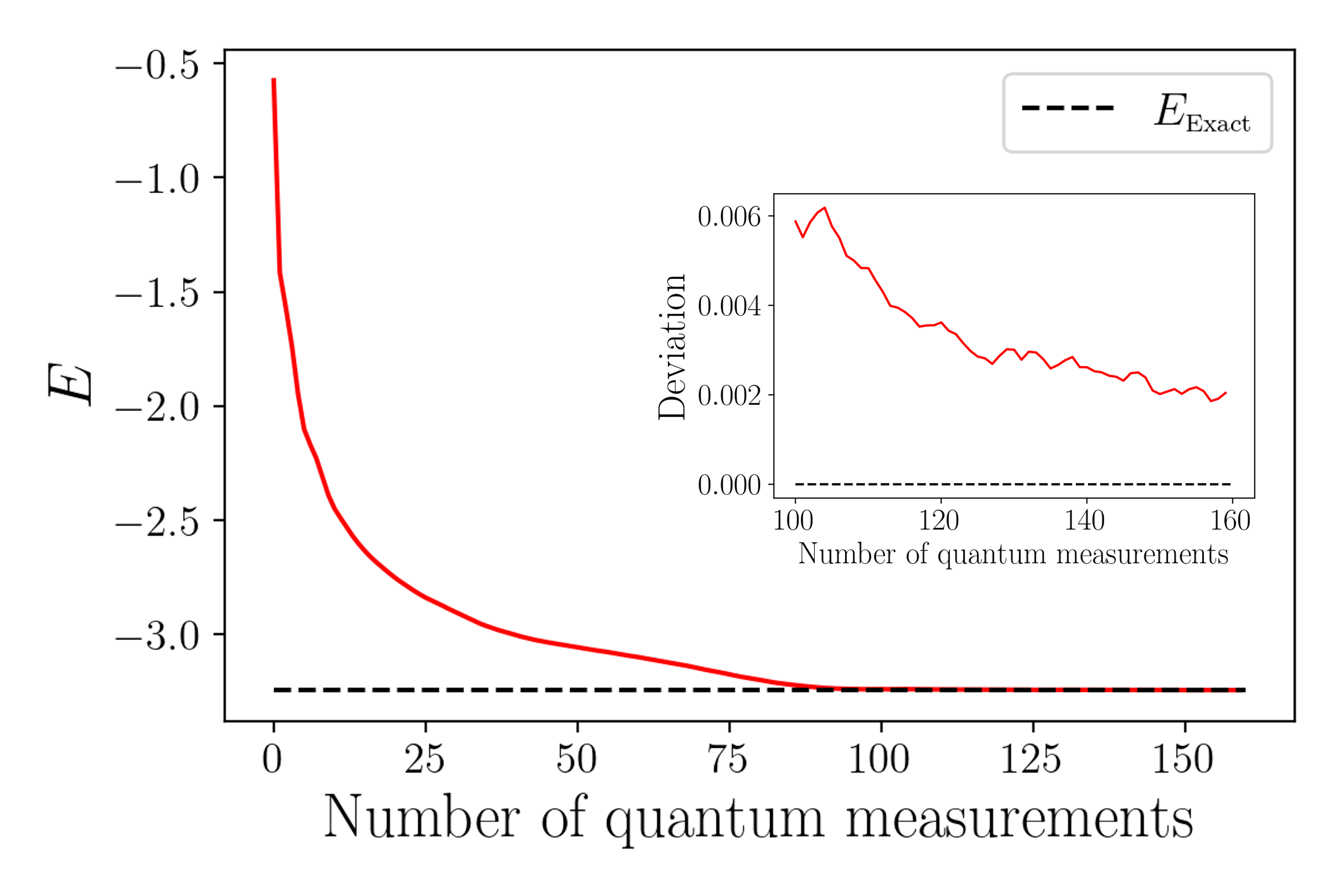}
\caption {\label{Fig2} (Color online) Eigenvalue $E$​ as a function of the number of quantum measurements performed 
by the quantum eigensolver (QE) on the D-Wave quantum annealer for H$_6$ chain. Inset: the derivation observed by 
discrete binary samplings of D-Wave quantum annealer. Here, the derivation arising in the QE method is corrected 
using the Lanczos algorithm, where the converged QE result is employed as the initial state of Lanczos method.}
\end{figure}

The remainder of this paper is organized as follows. Section~\ref{GBBD} describes the detailed techniques 
of the GBBD method, including graph decomposition and Hamiltonian reduction. In Section~\ref{Results}, we 
present and analyze the computational results of the H$_2$ chain systems. Finally, our conclusions 
are summarized in Section~\ref{Conclusion}.

\section{Graph-based Block-Diagonalization method\label{GBBD}}

We obtained the FCI Hamiltonian of molecular systems using 
the OpenFermion package~\cite{McClean2017}. To reduce computational cost, we restricted the Hilbert space to a 
subspace containing only configurations with a fixed $m$. The resulting Hamiltonian in this subspace has a size 
of $n \choose m$. To efficiently compute the $E_N$​ of this large and 
sparse Hamiltonian, we introduced the GBBD method, which is based on graph decomposition into connected components. 
In this approach, each basis state $\vert X \rangle$ is represented as a node in a graph, and non-zero matrix 
elements of the Hamiltonian define the edges between these nodes. The corresponding objective function on the graph 
is given by
\begin{equation}
f(x_0,x_1,...,x_L) = \langle X \vert \hat{H} \vert X \rangle,
\end{equation}
where $\vert X \rangle = [x_0, x_1, \dots, x_L]$ denotes a basis configuration. Due to the extreme sparsity of the 
Hamiltonian, the resulting graph is not fully connected. Using the networkx.connected components method from the 
NetworkX Python library~\cite{Hagberg2008}, we identified disconnected subgraphs that correspond to independent 
submatrices of the Hamiltonian. This allowed us to reconstruct a block-diagonalized form, where each block (or 
cluster) can be treated independently and varies in size. We confirmed that the eigenvalues of the ground state and 
the first excited state were obtained from the largest and second-largest block matrices, respectively. The 
eigenvalues of higher excited states were also computed, either from smaller blocks or by solving a modified 
Hamiltonian constrained by orthonormality conditions with respect to the previously obtained lower-energy 
eigenstates.

\section{Results\label{Results}}

As a benchmark system, we selected linear H chains ranging from H$_2$ to H$_{12}$.
The H chain model has been studied extensively in the past~\cite{Kadzielawa2015, Motta2020}. 
The block-diagonalized 
Hamiltonian were constructed using the GBBD approach with the reduced Hamiltonian. In Fig.~\ref{Fig1} we plotted the 
logarithmic sizes of four types of Hamiltonian as a function of the number of H$_2$: (i) the full Hamiltonian in 
the Fock space, (ii) the reduced Hamiltonian with a fixed number $m$ of electrons, (iii) the block matrix 
corresponding to the energy level $E_0$​, and (iv) the block matrix corresponding to the energy level $E_1$. ‘GSC’ and 
‘1$^{\text{st}}$’ denote the ground-state configuration and the first excited-state configuration within 
the block-diagonalized subspaces, respectively. As shown in Fig.~\ref{Fig1}, we found that the sizes of the block 
matrices corresponding to $E_0$ and $E_1$ increase exponentially with the number of H$_2$, similar to the full 
Hamiltonian in Fock space, although they remain significantly smaller than the full Hamiltonian itself. In addition, 
the block matrices corresponding to $E_0$ and $E_1$ are highly sparse.

\begin{figure}
\includegraphics[width=1.0\columnwidth]{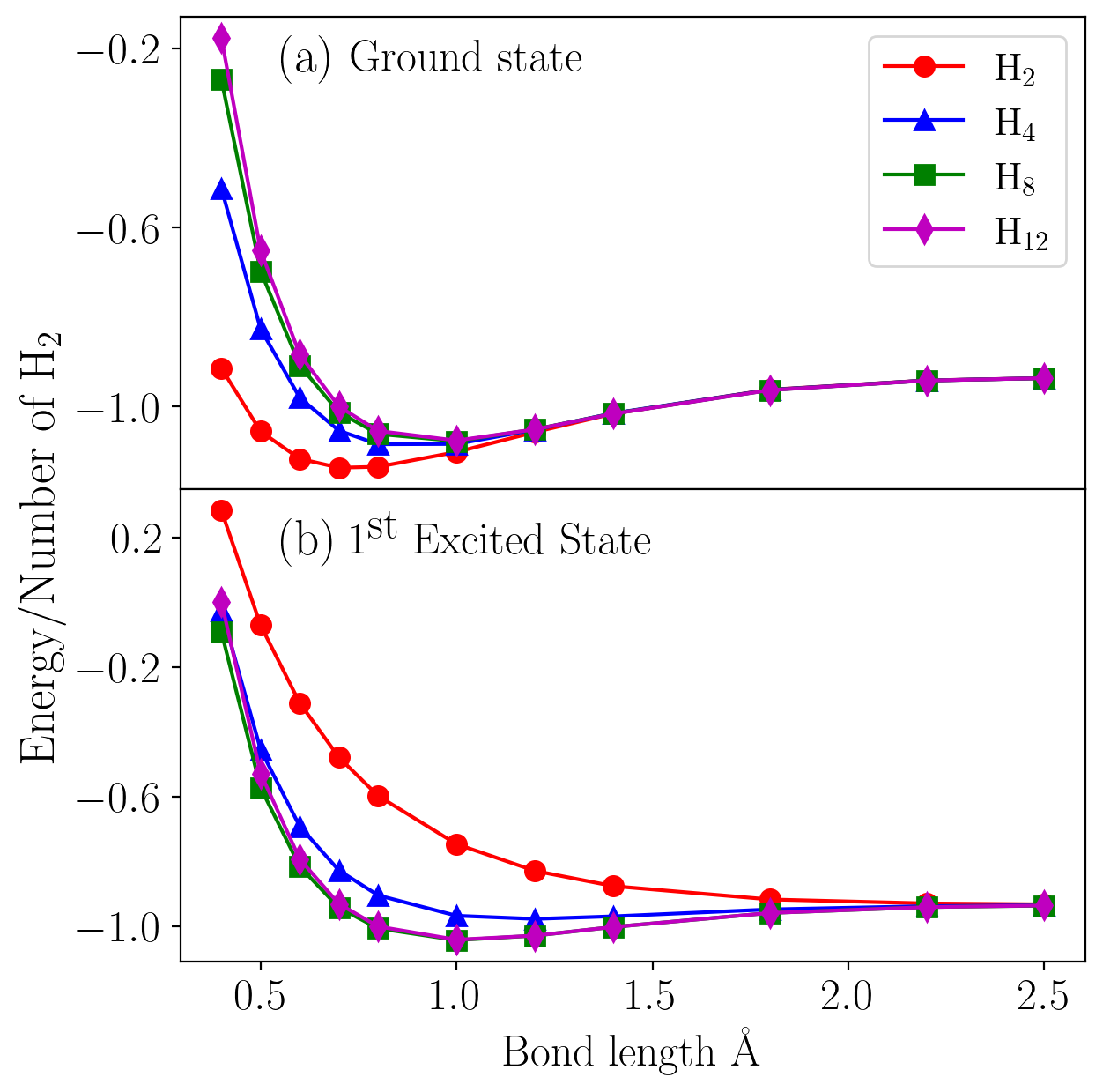}
\caption {\label{Fig3} (Color online) (a) The ground-state eigenvalue and (b) the first-excited state eigenvalue per 
the number of H$_2$ molecule as a function of the bonding length. The unit of the bond length is angstrom 
$\mathrm{\AA}$.}
\end{figure}

We computed the exact eigenvalues of the block matrix corresponding to the energy level $E_0$ using the QE method 
with iterative measurements on the D-Wave quantum annealer~\cite{Park2024}. The QE algorithm exhibits linear scaling 
$L$ with respect to 
the matrix size, demonstrating strong potential as a practical eigensolver in conjunction with the continued 
advancement of quantum hardware. Fig.~\ref{Fig2} shows the $E_0$ as a function of the number of quantum measurements 
performed on the D-Wave device for H$_6$ molecule. Approximately $100$ measurements were required to achieve 
convergence. The error due 
to discrete binary sampling, shown in the inset of Fig.~\ref{Fig2}, remained within $10^{-3}$. To further reduce the 
error, we employed the Lanczos method using the converged QE result as the initial state of the Lanczos approach.

\begin{figure}
\includegraphics[width=1.0\columnwidth]{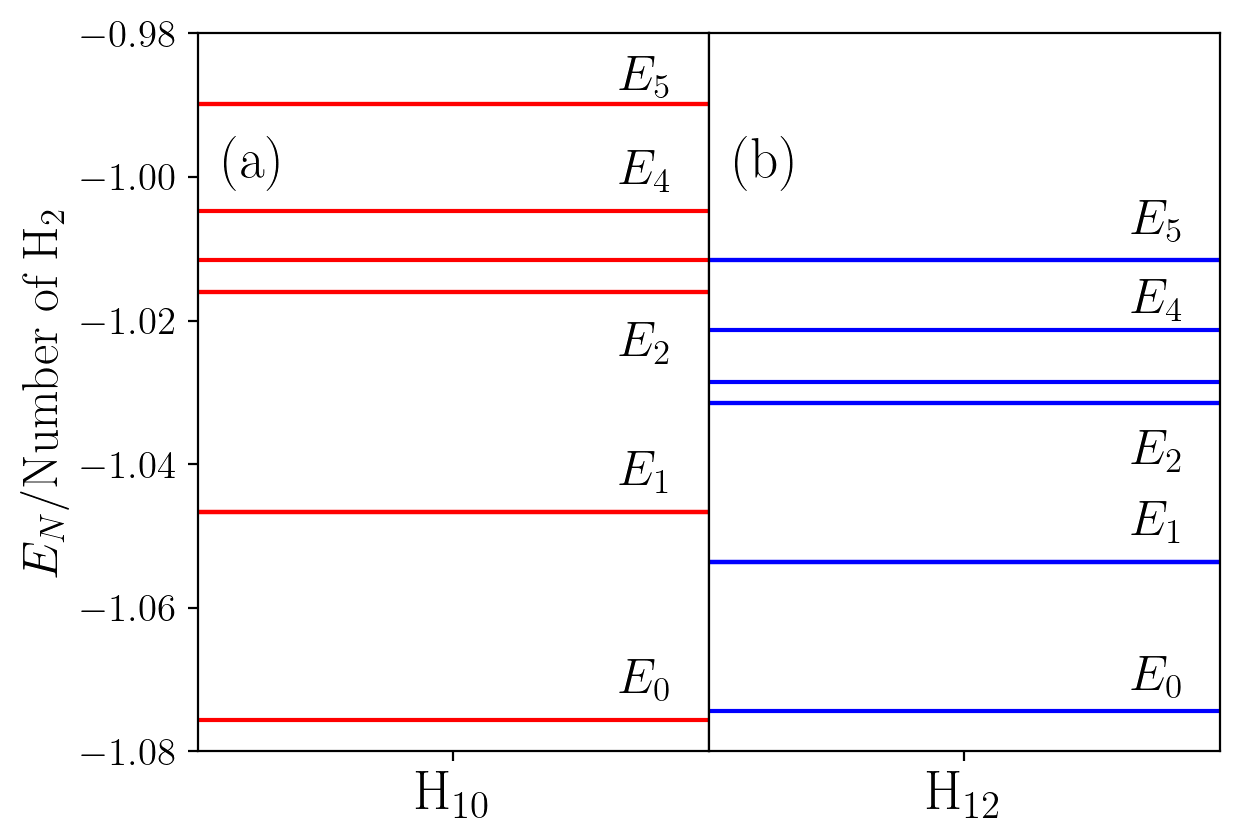}
\caption {\label{Fig4} (Color online) Several eigenvalues $E_N$ per the number of H$_2$ molecule for (a) H$_{10}$ and 
(b) H$_{12}$ with an optimal bond length of 0.9 $\mathrm{\AA}$ in the low-energy sector. $N$ means the energy level.}
\end{figure}

Next, we present the $E_0$ and  $E_1$ per the number of H$_2$ molecule as a function of the bond length in 
Fig.~\ref{Fig3}(a) 
and (b), respectively. The unit of the bond length is angstrom $\mathrm{\AA}$. We confirmed that the values of $E_0$ 
and $E_1$, corresponding to the eigenvalues of the largest and second-largest block matrices, are in agreement with 
those obtained from the full subspace Hamiltonian of size ${n \choose m}$. This result indicates that the dominant 
low-energy eigenstates are localized within the largest connected components of the reduced configuration space. We 
also found that the optimal bonding length corresponding to the minimum $E_0$ is approximately $0.9$
$\mathrm{\AA}$ for all H chains. At large bonding lengths, the energy per the number of H$_2$ molecule 
converges to the same value, reflecting the complete separation of the H atoms.

We computed the additional excited-state eigenvalues obtained from other blocks or modified Hamiltonian in 
each block matrix. Here, the form of the modified Hamiltonian of block matrix $\hat{H}^{\text{Block}}$ is given as
\begin{equation}
\hat{H}^{\text{Mod}} = \hat{H}^{\text{Block}} + w \vert \Psi \rangle \langle\Psi \vert, 
\end{equation}
where $\vert \Psi \rangle$ is a previously computed eigenstate and $w$ is a penalty weight. This shift 
effectively suppresses previously computed eigenstates, allowing the extraction of higher excited states 
through iterative diagonalization. We plotted several eigenvalues per the number of H$_2$ molecule for H$_{10}$ and 
H$_{12}$ with optimal bonding length $0.9$ $\mathrm{\AA}$ at the low-energy sector in Fig.~\ref{Fig4}(a) and (b), respectively. In 
addition, the eigenvalues of H$_{10}$ were compared with the exact results and showed excellent agreement.
The eigenvalues for H$_{12}$ could not be compared with the exact ones due to memory limitations caused by the 
massive Hamiltonian matrix on our laptop. We observed that the energy gaps between several eigenvalues in H$_{12}$ 
are more compressed than those in H$_{10}$.

\begin{figure}
\includegraphics[width=1.0\columnwidth]{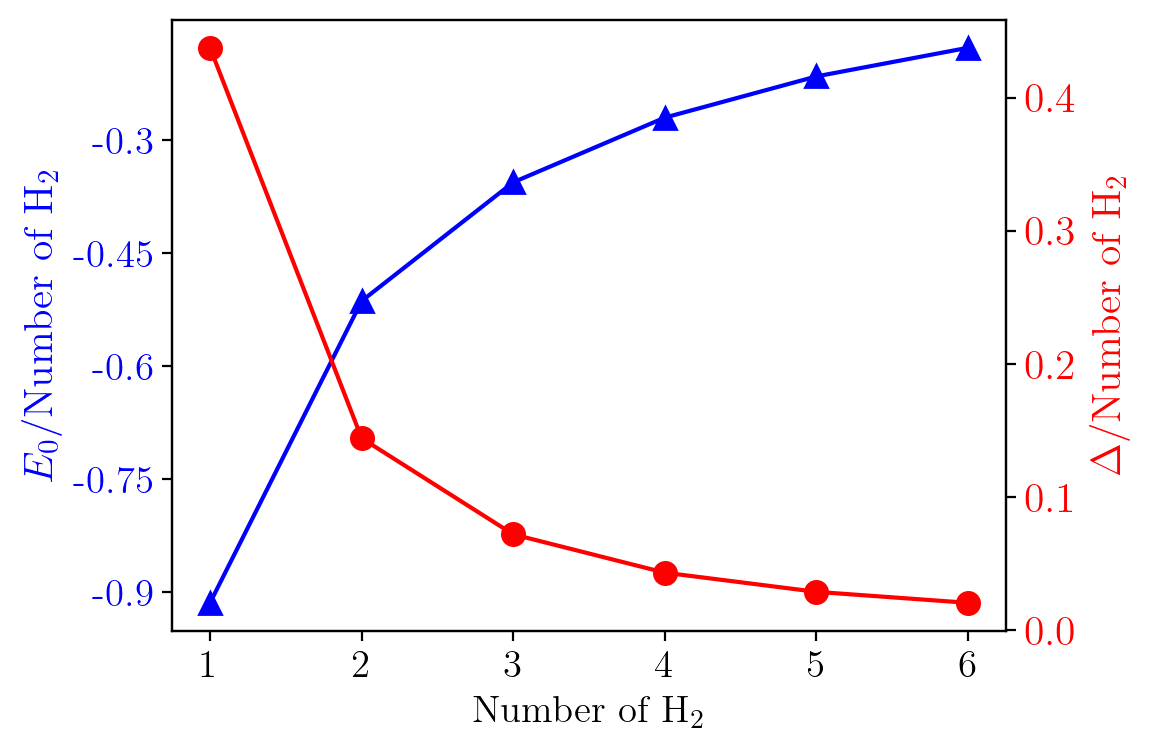}
\caption {\label{Fig5} (Color online) (Left y-axis): Ground-state energy $E_0$; (right y-axis): energy gap $\Delta = 
E_0 - E_1$, both plotted as functions of the number of H$_2$ molecule at optimal bond length 0.9 $\mathrm{\AA}$.}
\end{figure}

Finally, we analyzed the $E_0$ and the energy gap $\Delta = E_0 - E_1$ as functions of the number of H$_2$ units at 
0.9 $\mathrm{\AA}$. The results are presented in Fig.~\ref{Fig5}. We found that the value of $E_0$ per the number of 
H$_2$ molecule increases as the number of H$_2$ molecules increases, indicating that longer H$_2$ chains are less stable, and that 
the isolated H$_2$ molecule is the most stable configuration. Furthermore, the $\Delta$ decreases as the number of 
H$_2$ molecules increases, suggesting a tendency toward gap closure. In the limit of an infinite chain, $\Delta$ may 
converge to zero, implying a metallic behavior.

\section{Conclusion\label{Conclusion}}
We developed the GBBD approach for the FCI Hamiltonian of molecular systems to efficiently compute the exact 
eigenvalues of 
low-energy states. The molecular Hamiltonian were constructed in Fock space using the OpenFermion package~\cite{McClean2017}. We 
restricted the basis to bitstring states with a fixed $m$, 
reducing the Hilbert space size to $n \choose m$ configurations. Within this subspace, non-zero matrix elements of 
the Hamiltonian were mapped to edges in a graph, which naturally decomposes into disconnected clusters. Each cluster 
corresponds to an independent block in the block-diagonalized Hamiltonian. The eigenvalues in the low-energy sector 
were obtained by solving the eigenvalue problem for each block using a QE method on the D-Wave quantum 
annealer, supplemented by the Lanczos method for error correction. Notably, the ground-state and first excited-state 
eigenvalues were found to reside in the largest and second-largest blocks, respectively. Higher excited-state 
eigenvalues were also computed from other blocks or by solving modified Hamiltonians constrained by orthonormality 
conditions on the previously obtained eigenstates. We applied the GBBD method to linear H$_2$ chains ranging from 
H$_2$​ to H$_{12}$​, and the results showed excellent agreement with exact diagonalization, validating the accuracy and 
efficiency of our approach.

\section{Acknowledgements\label{Ack}}
This work was supported by Institute of Information and communications Technology Planning Evaluation (IITP) grant 
funded by the Korean government (MSIT) (No. RS-2023-0022952422282052750001). We would like to thank June-Koo Kevin 
Rhee for valuable discussions.


\begin{thebibliography}{99}
\bibitem{Szabo1996} A. Szabo and N. S. Ostlund, Modern Quantum Chemistry: Introduction to Advanced Electronic 
Structure Theory, Dover Publications (1996).

\bibitem{Weisse2008} A. Weiße and H. Fehske, Exact Diagonalization Techniques, Computational Many-Particle 
Physics Lecture Notes in Physics (2008).

\bibitem{Knowles1984} P. J. Knowles and N. C. Handy, A new determinant-based full configuration interaction 
method, Chem. Phys. Lett. {\bf 111}, 315-321 (1984).

\bibitem{Peruzzo2014} A. Peruzzo, J. McClean, P. Shadbolt,  M.-H. Yung, X.-Q. Zhou, P. J. Love, 
A. A.-Guzik and J. L. O’Brien, A variational eigenvalue solver on a photonic quantum processor, Nat. 
Commun. {\bf 5}, 4213 (2014).

\bibitem{Arute2020} F. Arute, K. Arya, R. Babbush, et al. (Google AI Quantum and Collaborators), Hartree-Fock on a 
superconducting qubit quantum computer, Science {\bf 369}, 1084-1089 (2020).

\bibitem{Kandala2017} A. Kandala, A. Mezzacapo, K. Temme, M. Takita, M. Brink, J. M. Chow, 
and J. M. Gambetta, Hardware-efficient variational quantum eigensolver for small molecules and quantum magnets, 
Nature {\bf 549}, 242-246 (2017).

\bibitem{McArdle2020} S. McArdle, S. Endo, A. A.-Guzik, S. C. Benjamin, and X. Yuan,
Quantum computational chemistry, Rev. Mod. Phys. {\bf 92}, 015003 (2020).

\bibitem{White1992} S. R. White, Density Matrix Formulation for Quantum Renormalization Groups, Phys. Rev. Lett. 
{\bf 69}, 2863 (1992).

\bibitem{White1993} S. R. White and R. M. Noack, Real-space quantum renormalization groups, Phys. Rev. Lett. {\bf 
68}, 3487 (1992).

\bibitem{Chan2011} G. K.-L. Chan and S. Sharma, The Density Matrix Renormalization Group in Quantum 
Chemistry, Annu. Rev. Phys. Chem. {\bf 62}, 465-481 (2011).

\bibitem{Baiardi2020} A. Baiardi and M. Reiher, The density matrix renormalization group in chemistry and 
molecular physics: Recent developments and new challenges, J. Chem. Phys. {\bf 152}, 040903 (2020).

\bibitem{Foulkes2001} W. M. C. Foulkes, L. M. Mitas, R. J. Needs, and G. Rajagopal, Quantum Monte Carlo simulations 
of solids, Rev. Mod. Phys. {\bf 73}, 33 (2001).

\bibitem{Ceperley1980} D. M. Ceperley and B. J. Alder, Ground State of the Electron Gas by a Stochastic Method, Phys. 
Rev. Lett. {\bf 45}, 566 (1980).

\bibitem{Booth2009} G. H. Booth, A. J. W. Thom, and A. Alavi, Fermion Monte Carlo without fixed nodes: A game 
of life, death, and annihilation in Slater determinant space, J. Chem. Phys. {\bf 131}, 054106 (2009).

\bibitem{Austin2012} B. M. Austin, D. Y. Zubarev, and W. A. Lester, Quantum Monte Carlo and related approaches, Chem. 
Rev. {\bf 112}, 263-288 (2012).

\bibitem{Bauer2020} B. Bauer, S. Bravyi, M. Motta, and G. K.-L. Chan, Quantum Algorithms for Quantum Chemistry and 
Quantum Materials Science, Chem. Rev. {\bf 120}, 12685-12717 (2020).

\bibitem{Qunova2025} A. P.-Jarman, S. McFarthing, D. H. Kang, P. Yoo, E. E. Elala, R. P.-Jarman, P. M. Nakliang, J. 
Kim and J.-K. K. Rhee, HIVQE: Handover Iterative Variational Quantum Eigensolver for Efficient Quantum Chemistry 
Calculations, arXiv:2503.06292 (2025).

\bibitem{Bartlett2007} R. J. Bartlett and M. Musiał, Coupled-cluster theory in quantum chemistry, Rev. Mod. Phys. 
{\bf 79}, 291 (2007).

\bibitem{Sherrill1999} C. D. Sherrill and H. F. Schaefer III, The Configuration Interaction Method: Advances in 
Highly Correlated Approaches, Adv. Quantum Chem. {\bf 34}, 143-269 (1999).

\bibitem{McClean2017} J. R. McClean, K. J. Sung, et al., {OpenFermion}: The Electronic Structure Package for Quantum 
Computers, arXiv:1710.07629 (2017).

\bibitem{Park2024} H. Park and H. Lee, Quantum eigensolver on extension of optimized binary configurations, Phys. 
Rev. B {\bf 110}, 205142 (2024).

\bibitem{Hagberg2008} A. A. Hagberg, D. A. Schult and P. J. Swart, Exploring network structure, dynamics, and 
function using NetworkX, in Proceedings of the 7th Python in Science Conference (SciPy2008) (2008).

\bibitem{Kadzielawa2015} A. P. Kadzielawa, A. Biborski and J. Spa\l{}ek, Discontinuous transition of 
molecular-hydrogen chain to the quasiatomic state: Combined exact diagonalization and ab initio approach, Phys. Rev. 
B {\bf 92}, 161101 (2015).

\bibitem{Motta2020} M. Motta, C. Genovese, F. Ma, Z.-H. Cui, R. Sawaya, G. K.-L. Chan, N. Chepiga, P. Helms, C. 
Jim\'enez-Hoyos, A. Millis, U. Ray, E. Ronca, H. Shi, S. Sorella, E.M. Stoudenmire, S. White, S. Zhang, Ground-State 
Properties of the Hydrogen Chain: Dimerization, Insulator-to-Metal Transition, and Magnetic Phases, Phys. Rev. X {\bf 
10}, 031058 (2020).

\end{thebibliography}
\end{document}